\documentclass[12pt]{article}
\usepackage{graphicx}
\usepackage{enumerate} 
\def\lsim{\mathrel{\vcenter{\hbox{$<$}\nointerlineskip\hbox{$\sim$}}}}
\newcommand{\be}{\begin{equation}}
\newcommand{\ee}{\end{equation}}
\newcommand{\ba}{\begin{eqnarray}}
\newcommand{\ea}{\end{eqnarray}}

\def\21{$SU(2) \otimes U(1) $}

\def\lsim{\raise0.3ex\hbox{$\;<$\kern-0.75em\raise-1.1ex\hbox{$\sim\;$}}}
\def\gsim{\raise0.3ex\hbox{$\;>$\kern-0.75em\raise-1.1ex\hbox{$\sim\;$}}} 

\newcommand{\mx}{\left[\begin{array}}
\newcommand{\finmx}{\end{array}\right]} 
\newcommand{\mxp}{\left(\begin{array}} 
\newcommand{\finmxp}{\end{array}\right)} 
\def\beq{\begin{equation}}
\def\eeq{\end{equation}}
\def\bea{\begin{eqnarray}}
\def\eea{\end{eqnarray}}

\def\mathbf#1{\hbox{\bf #1}}
\def\textrm#1{\hbox{#1}}

\def\lsim{\raise0.3ex\hbox{$\;<$\kern-0.75em\raise-1.1ex\hbox{$\sim\;$}}}
\def\gsim{\raise0.3ex\hbox{$\;>$\kern-0.75em\raise-1.1ex\hbox{$\sim\;$}}}
\newcommand {\ignore}[1]{}

\begin{document}
\vspace*{-1in}
\renewcommand{\thefootnote}{\fnsymbol{footnote}}
\begin{flushright}
\texttt{
} 
\end{flushright}
\vskip 5pt
\begin{center}
{\Large{\bf Cosmo MSW effect for mass varying neutrinos}}
\vskip 25pt 

{\sf Pham Quang Hung$^1$, Heinrich P\"as$^2$}
\vskip 10pt
{\small $^1$ 
\it Department of Physics, 
University of Virginia,\\ 382 McCormick Road, Charlottesville
VA 22904-4714, USA}\\
{\small $^2$ \it Institut f\"ur Theoretische Physik und Astrophysik,
Universit\"at W\"urzburg,\\ D-97074 W\"urzburg, Germany}\\
\vskip 10pt

{\bf Abstract}
\end{center}

\begin{quotation}
{\small 
We consider neutrinos with varying masses which arise in scenarios relating
neutrino masses to the dark energy density in the universe.
We point out that the neutrino mass variation can lead to level crossing
and thus a cosmo MSW effect, having dramatic consequences for
the flavor ratio of astrophysical neutrinos. 
}
\end{quotation}

\vskip 20pt  

\setcounter{footnote}{0}
\renewcommand{\thefootnote}{\arabic{footnote}}

Two of the most mysterious puzzles in particle physics and cosmology 
concern the mechanism generating small neutrino masses and the origin
of the dark energy in the universe. There exists, however, an amazing
coincidence between the order of magnitude of neutrino masses, 
$m_{\nu}< 1$~eV
and the vacuum density being responsible for the dark energy,
$\rho_V \approx (10^{-3}$~eV$)^4$. 
In ref. \cite{Hung:2000yg} (compare also \cite{Singh:1994nt}) 
this coincidence was explained by 
generating sterile neutrino masses of order $10^{-3}$~eV
via interactions with a scalar field, 
the ``acceleron''. Recently a more elaborated scenario   
\cite{Fardon:2003eh} was published, where the sterile neutrino
mass variation is transmitted to the active sector in a seesaw framework,
when integrating out the singlets under the Standard Model gauge group.
This scenario conserves the ratio of the neutrino and dark energy contributions
to the total energy density of the universe and
implies the sterile neutrino mass term $M_s$ 
to decrease with the 
cosmic evolution, while the effective active 
neutrino masses $m_{\nu}$ vary like the inverse neutrino density,
\ba{}
m_{\nu}(z) &=& (1+z)^{3 w} m_{\nu 0}
\rightarrow (1+z)^{-3} m_{\nu 0}, \\
M_s (z) &\propto& m_{\nu}^{-1} (z) \rightarrow (1+z)^{3} M_{s 0}.
\label{massvar}
\ea
Here $z$ is the cosmological redshift, $w \approx -1$ 
defines the equation of state
of the dark energy, and the neutrinos are
assumed to propagate in a  non-relativistic background. 

Several effects to test this hypothesis
have been discussed in \cite{Fardon:2003eh}, including observation of 
sterile neutrino states being in conflict with big bang nucleosynthesis, 
conflicts of terrestrial and astronomical neutrino mass measurements,
as well as the predicted relation of the dark energy equation of state
and the cosmological neutrino mass. Implications for baryogenesis via 
leptogenesis in mass varying neutrino scenarios have been discussed in
\cite{Bi:2003yr}. For more recent work see also \cite{recent}.

Here we focus on a spectacular effect which seems to be overlooked so far, 
namely the possibility of an MSW effect for cosmological
neutrinos {\it in vacuo}, 
which is possible if the variation of
neutrino masses leads to level crossing of the associated mass eigenstates
(for a similar discussion in a different context, see 
\cite{Stephenson:1996cz}).
Such level-crossings appear naturally in a large class of models. An obvious 
possibility is to assume active states with constant masses
being only weakly coupled to 
mass varying sterile states, as would be the case in \cite{Hung:2000yg}.
Another possibility is a seesaw framework as in
\cite{Fardon:2003eh}, with some of the singlet states being light and 
only weakly mixed
with the active states, leading to a mass matrix of the kind
\be{}
{\cal M} \sim \left(\begin{array}{ccc}
0 & m_D & \epsilon \\ m_D & M_s & 0 \\ \epsilon & 0 & m_s  \end{array}\right),
\label{massmatrix}
\ee   
with
$\epsilon \ll m_D \lsim m_s \ll M_s$, in the early universe.
Here $m_D$ denotes the Dirac mass and $m_s$, $M_s$ are Majorana masses of the
light and heavy singlets, respectively.
Integrating out $M_s$ yields a mass matrix 
\be{}
{\cal M} \sim \left(\begin{array}{cc}
-m_D^2/M_s & \epsilon \\ \epsilon & m_s  \end{array}\right).
\ee 
Assuming now the masses of the sterile states to decrease, i.e. $m_s$, 
$M_s\rightarrow 0$ while keeping $m_s/M_s$ constant, 
the mass eigenstates experience active-sterile 
level crossing at a resonance point with $-m_D^2/M_s = m_s$.
This scheme is motivated by assuming the Majorana masses to be generated 
by couplings to the same
singlet acceleron field, while the Dirac masses originate from couplings to 
the Standard Model Higgs. The evolution of all three mass eigenstates
$\tilde{m}_i$ with $M_s$ in such a scenario is illustrated in Fig.~1.
Further possibilities to generate level crossing 
involve more complicated flavor structures, allowing the individual flavors
to evolve with different time-dependence.

\begin{figure}
\centerline{\resizebox{12cm}{7cm}{\includegraphics{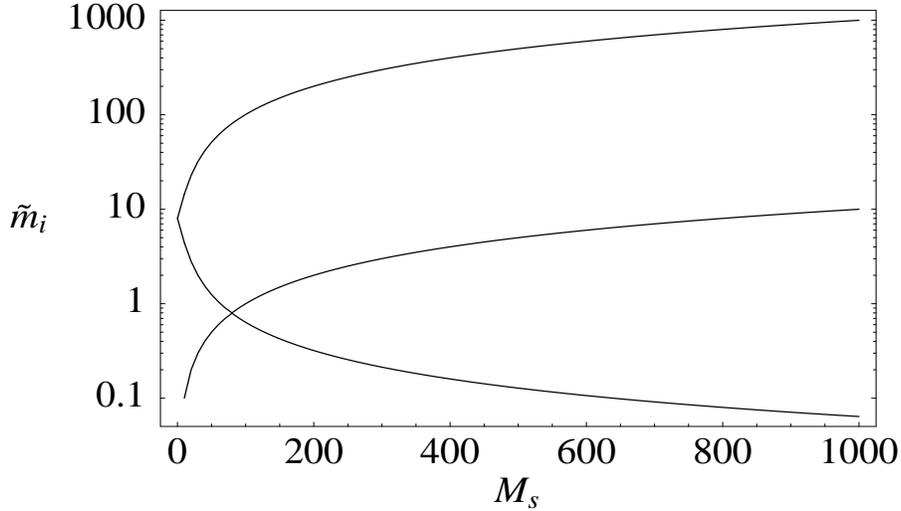}}}
 \caption{Level crossing in the seesaw scheme for mass varying neutrinos
-- schematically. Shown is the evolution of the three mass eigenstates
$\tilde{m}_i$ in a scenario described by the mass matrix 
(\ref{massmatrix}).
}
 \label{fig:levelcrossing}
\end{figure}

The two illustrative examples discussed above can both be described with
one sterile neutrino flavor state $\nu_s$
possessing a mass term varying
according to (\ref{massvar}) and an active state $\nu_a$
with a constant 
mass \footnote{In the seesaw case the varying sterile mass dominates the 
mass squared difference to the active state.}.
Generalization to different scenarios is straightforward, 
since a flavor-blind potential doesn't enter the effective mixing angles
and oscillation probabilities, and thus each logical possibility can be 
reduced to the chosen option.

The evolution equation in flavor space reads \cite{msw,moha}
\be{}
i\frac{d}{dt}\left(\begin{array}{c} \nu_a(t)\\ 
\nu_s(t)  \end{array}\right) = \tilde{H} \left(\begin{array}{c} \nu_a(t)\\ 
\nu_s(t)  \end{array}\right),
\ee
where 
\ba{}
\tilde{H}
 &=&E + \frac{m_1^2 + m_2^2}{4E} + 
\left(\begin{array}{cc}   
(- \frac{\delta m^2}{4E} \cos 2 \theta 
+ \frac{\delta m_{\rm cosm}^2}{2E}) & \frac{\delta m^2}{4E} \sin 2 \theta \\
  \frac{\delta m^2}{4E} \sin 2 \theta & \frac{\delta m^2}{4E} \cos 2 \theta 
\end{array}\right). 
\label{hamilt}
\ea
Here $m_i$ and $E$ denote the neutrino mass eigenstates and energy, 
respectively, and the mixing angle 
\be{}
\theta = \frac{1}{2}\arcsin \left(\frac{\epsilon}{m_1-m_2}\right) 
\ee
parametrizes the $z=0$ 
neutrino mixing matrix. 

The mass squared differences as measured in 
present experiments and
due to the cosmological mass variation (\ref{massvar}) are
$\delta m^2= |m_2^2 - m_1^2|$ and 
\be{}
\delta m_{\rm cosm}^2=m_{s 0}^2 \left[(1+z)^6 - 1\right],
\label{dmsqcos}
\ee
respectively, 
assuming $w=-1$. 
The Hamiltonian $\tilde{H}$ is diagonalized by the effective 
mixing angle
\be{}
\tan 2 \tilde{\theta} = \frac{\delta m^2 \sin 2 \theta}
{\delta m^2 \cos 2 \theta - \delta m_{\rm cosm}^2}. 
\ee
The mass eigenvalues are given by
\be{}
E_{\alpha}= E + \frac{\tilde{m}_\alpha^2}{2 E},
\ee
where
\be{}
\tilde{m}^2_{1,2}=\frac{1}{2}\left[
\left(m_1^2 + m_2^2 + \delta m^2_{\rm cosm}\right)
\mp \sqrt{(\delta m^2 \cos 2 \theta - \delta m^2_{\rm cosm})^2
+ \delta m^2
\sin^2 2 \theta}\right],
\ee
and the resonance occurs for
\be{}
\delta m^2_{\rm cosm} = \delta m^2 \cos 2 \theta.
\ee

For an adiabatic transition, the adiabaticity parameter,
evaluated at the resonance, has to be large,
\be{}
\gamma = \frac{\delta m^2 \sin^2 2 \theta}{E \cos 2 \theta}
\left| \frac{H_0}{\delta m^2_{\rm cosm}}~f(z)
~\frac{d(\delta m^2_{\rm cosm})}{dz} 
\right|^{-1}_{z_{\rm res}} \gg 1,
\label{hubble}
\ee
where the Hubble relation, $z=H_0~ f(z)~ d$ with
\be{}
f(z)=\sqrt{\Omega_M (1+z)^3 + \Omega_{\Lambda}}
\ee
for a flat universe
has been assumed \cite{Kolb:vq}. The cosmological parameters are given by
$H_0=70$~km/(s~Mpc), $\Omega_{\Lambda}= 0.73$ and $\Omega_M=0.27$ 
(see, e.g. \cite{Spergel:2003cb}), so that
$f(z)\simeq 1$ for small $z \ll 1$ and $f(z)\simeq 0.3~z^{3/2}$ for large 
$z \gg 1$.
Since $H_0\simeq 10^{-33}$~eV, adiabaticity is easily fulfilled, even 
for PeV neutrinos.
In this case, the Landau-Zener-St\"uckelberg probability 
$P^{LZS}=\exp\left(-\frac{\pi}{4}\gamma\right)$
vanishes, and the
oscillation probability is given by
\be{}
P(\nu_a \rightarrow \nu_s)=
\frac{1}{2}(1-\cos 2 \theta \cos 2 \tilde{\theta}).
\ee
It is interesting to note, that, contrary to the common MSW effect, the 
cosmo MSW effect depends via (\ref{dmsqcos}) on both $z$ and $m_{\nu 0}$,
and thus exhibits information on absolute neutrino masses.

In the following we calculate the oscillation probabilities for 
neutrinos from distant astrophysical sources. For this purpose we 
generalize the two-neutrino framework to a 3+1 generation framework,
by assuming $m_D^2/M_s$ to be a $3 \times 3$ matrix. The mixing 
$\sin \theta \sim 0.1$  and mass squared difference
$\delta m^2 \sim 0.1$~eV$^2$ are chosen, as can be assumed in 3+2 models
\cite{Sorel:2003hf}
in order to accomodate the LSND result \cite{lsnd}, and
$m_{s 0}\approx 0.1$~eV is assumed. In this case the active states
are degenerated with masses at a scale $\sim \sqrt{m_{s0}^2+\delta m^2}\sim
0.45$~eV \footnote{Note, that cosmological neutrino mass bounds do
not apply in the mass varying neutrino scenario.}, 
in accordance with the recently claimed evidence for neutrinoless 
double beta decay \cite{Klapdor-Kleingrothaus:2004wj}, and the 
heavy mass $M_{s0}\sim 20$~eV could 
play the role of the fifth state \cite{Sorel:2003hf}.

While local neutrino sources such as a supernova in the Large Magellanic Cloud
(LMC)
are not significantly affected by cosmological neutrino mass variation, 
neutrino telescopes may be sensitive
to neutrinos from active galactic nuclei (AGN's) at distances of 
1000~Mpc ($z=0.3$) and energies of a PeV
\cite{Halzen:1998be}.
The corresponding effect on the absolute neutrino mass is
\ba{}
\left(\frac{m_{s}(z)}{m_{s 0}} \right)^2  
&=& 4.8~~~{\rm (AGN)} 
\ea
and we obtain
\be{}
P(\nu_a \rightarrow \nu_s)=0.96
~~~{\rm (AGN)}.
\ee
It is obvious that the
flavor ratios of astrophysical neutrino fluxes obtained may significantly
deviate from the expected \cite{Ahluwalia:2001xc}
$\nu_e:\nu_\mu:\nu_\tau:\nu_s$
ratio of 1:1:1:0  after decoherence of flavor into mass eigenstates. 
For normal hierarchical neutrinos
after the first level crossing the 
information about the $\tau$ neutrino flux is lost and
the initial flavor spectrum $1:2:0:0$ 
is transformed resonantly into the characteristic
$1:\frac{1}{2}:\frac{1}{2}:1$\footnote{These flavor ratios have
also been derived in \cite{pakv}.}. 
Such a flavor ratio corresponds to a muon 
to shower ratio of about
5 in next generation neutrino telescopes such as IceCube \cite{Beacom:2003nh}. 
For inverse hierarchical neutrinos, the
electron neutrinos become sterile and disappear, and the resulting flavor ratio
is 0.3:0.85:0.85:1, corresponding to a muon to shower ratio of 20
\cite{Beacom:2003nh}. These results do not depend on whether the resonance 
was reached before or after decoherence.
While the standard MSW resonance depends on the beam 
energy via the adiabaticity condition (\ref{hubble}), the effect here is 
essentially energy 
independent due to the enhancement of the adiabaticity parameter $\gamma$ 
by $H_0^{-1}$.

As has been mentioned above, local astrophysical 
sources such as SN87A in the LMC would not exhibit this effect. 
Such a distance dependent 
characteristic flavor composition would provide a strong evidence for
neutrino mass variation. It also could be a unique possibility to study 
the parameters triggering the neutrino mass evolution, such as the 
acceleron potential and the relic neutrino density. 
It should be stressed, though, that a clear signature
of this effect 
can be spoiled by neutrino overdensities at the source and, in non-standard
neutrino scenarios, also in
the galactic neighborhood, which fake the cosmological level crossing. 
However, since
typical neutrino source candidates such as 
AGN's bear large neutrino densities and the neutrino propagation in these 
backgrounds occurs on small time scales, while the neutrinos are extremely
high energetic, the adiabaticity condition,
\be{}
\gamma = \frac{\delta m^2 \sin^2 2 \theta}{E \cos 2 \theta}
\left| 
~\frac{d(\ln \delta m^2_{\rm cosm})}{dt} 
\right|^{-1}_{t_{\rm res}} \gg 1,
\ee
is not necessarily fulfilled. A conservative estimate assuming 
${\cal O}(\Delta \ln \delta m^2_{\rm cosm})\simeq 1$ and $\Delta t$ to be of 
the size of the Schwarzschild radius of a $10^9$ solar mass black hole,
$r_{\rm s} = 3 \cdot 10^{12}$~m,  results
in $1 < \gamma \simeq 40$ for $\delta m_{\rm atm}^2 =2.6 
\cdot 10^{-3}~{\rm eV}^2 < \delta m^2 < 0.1~{\rm eV}^2$ 
and PeV neutrino energies. Thus at least for
large neutrino energies in the multi-PeV region the process
is not adiabatic so that the 
resulting neutrino spectra are clearly distinguishable from 
level crossing due to the cosmological mass shift implied by the relic 
neutrino density. The non-adiabaticity is even stronger for a Gamma Ray Burst
source, where the time variability gives an estimate of the source size of 
about 1~light-second~$\simeq 10^9$~m.
Since these subtleties depend strongly on the 
astrophysical source, on assumptions about neutrino densities 
and on the scenario for 
mass variation chosen, a detailed discussion is beyond the scope of this work.
It should be kept in mind though, as a possible caveat.

In conclusion we discussed the effect of level-crossing in a 
mass-varying neutrino scenario. The resulting MSW effect in vacuo
is the same for neutrinos and antineutrinos, unlike the common MSW effect in 
matter. It 
can significantly distort the flavor ratios of
neutrino fluxes emitted
in active galactic nuclei, predicting characteristic muon to shower ratios
in next generation neutrino telescopes, which depend on the distance to the
source.

\section*{Acknowledgements}
We thank Michele Maltoni, Karl Mannheim and especially
Tom Weiler for useful discussions.
PQH is supported in parts by the US Department of Energy under grant 
No. DE-A505-89ER40518.
HP was supported by the Bundesministerium f\"ur Bildung
und Forschung (BMBF, Bonn Germany) under contract number 05HT1WWA2
and thanks the Instituto de Fisica Corpuscular (C.S.I.C.) Valencia and the 
Department of Physics of the University of Virginia for kind hospitality.

\end{document}